\documentclass[%
 reprint,
 amsmath,amssymb,
 aps,
prb,
]{revtex4-2}

\usepackage{graphicx}
\usepackage{dcolumn}
\usepackage{bm}
\usepackage{hyperref}
\usepackage{booktabs}

\begin{document}

    \title{%
        Spin-3/2 nuclear magnetic resonance:
        Exact solutions for aligned systems
        and implications for probing Fe-based superconductors%
    }

    \author{Jaafar N.\ Ansari}
    \email{jansari@gmu.edu}
    \affiliation{%
        Department of Physics and Astronomy, George Mason University,
        Fairfax, VA 22030, USA%
    }
    \author{Karen L.\ Sauer}
    \email{ksauer1@gmu.edu}
    \affiliation{%
        Department of Physics and Astronomy, George Mason University,
        Fairfax, VA 22030, USA%
    }
    \affiliation{%
        Quantum Science and Engineering Center, George Mason University,
        Fairfax, VA 22030, USA%
    } 

    \date{\today}

    \begin{abstract}
        The nuclear magnetic resonance (NMR) spectrum
        of spin-3/2 nuclei in a static magnetic field
        aligned with one of the electric field gradient (EFG)
        principal axes is developed analytically,
        based on fictitious spin-1/2 formalism.
        Compact closed-form expressions
        for the eigenstates and transitions frequencies,
        as well as the expectation value of the magnetic moment
        after resonant excitation, are derived.
        Emphasis is placed on defining and interpreting
        the associated Rabi frequencies,
        as a function of excitation direction and ellipticity.
        It is found transitions inherently fall into two subsets,
        depending on their sensitivity to excitation direction,
        with the Rabi frequency of one subset
        directly depending on the asymmetry of the EFG.  
        A natural application is the study of Fe-based superconductors,
        whose antiferromagnetic ordering at low temperatures
        leads to a strong intrinsic magnetic field
        aligned with the EFG principal axes.
        Zero external-field NMR spectra,
        from powder samples of two such Fe-based superconductors,
        BaFe$_2$As$_2$ and CaFe$_2$As$_2$,
        are analyzed and exemplify the simplicity
        in extracting the internal magnetic field,
        the quadrupole coupling constant, and the EFG asymmetry parameter,
        which are important for studying magnetic ordering,
        structural properties, phase transitions,
        and NMR dynamics.
        Results compare favorably to conventional high-field NMR experiments
        done with the rotation of single crystals.
        Overall, the physical insights,
        afforded by the exact and concise expressions,
        will lead to ready interpretation of spin-3/2 spectra
        as well as precipitate new experimental directions.
    \end{abstract}

    \maketitle

    \section{Introduction}
    Nuclei with spin greater than 1/2
    have both a magnetic and electric quadrupole moment. 
    The former couples to a magnetic field,
    the latter to an electric field gradient (EFG).
    Such nuclei can then serve as an internal probe
    of both magnetic and electric phenomena,
    and are addressable using the magnetic dipole's interaction
    with a radio-frequency excitation,
    i.e., nuclear magnetic resonance (NMR).
    Conventionally, experiments are often performed in two limits:
    (1) in a large magnetic field
    which is commonly known as high-field NMR,
    and (2) in the absence of a magnetic field,
    often referred to as nuclear quadrupole resonance (NQR).
    In this paper, we are concerned with the exact NMR theory of
    spin-3/2 nuclei in the presence of a magnetic field
    which is aligned with one of the EFG principal axes.
    The full continuum of results
    between the zero-field and high-field limits are addressed.

    We tackle this problem using spin-1/2 formalism,
    which has not been done before.
    Although the solution for this system is somewhat involved
    because the quadrupolar and Zeeman Hamiltonians do not commute,
    it has been solved exactly
    for arbitrary field directions
    \cite{muha1983, creel1983, bain2003, bain2017}
    and by perturbation theory many times over
    using a variety of techniques \cite{bain2012, dean1954}.
    In addition, software has been written
    which simulates such spectra \cite{perras2012, possa2011}.
    Previous work tackling the full solution
    is useful for computational analysis and modeling, however,
    they lack physical insight
    and the ability to be used for straightforward data analysis.
    With this in mind, a set of tractable analytical expressions
    in the EFG-aligned case is notably missing
    and would elevate the quality of both
    the quantitative and qualitative analysis of these systems.

    The present work addresses this problem,
    with the aim of going beyond
    just the eigenfrequencies, eigenstates, and transition rates,
    by deriving explicit expressions for the oscillating nuclear magnetization,
    including its direction,
    produced after a pulse of resonant radio-frequency excitation.
    The Rabi frequency, dictating the efficacy of the excitation,
    is proportional to the radio-frequency strength,
    as well as a modifying geometric term.
    This geometric term $ \lambda $ depends on the direction of the excitation field
    as well as the relative strengths of the static field components.
    An excitation pattern, specific to the transition frequency,
    can therefore be visualized through $ \lambda $.
    Such visualization has only been done in the pure NQR case~\cite{odin1999}. 
    Select transitions directly depend on the asymmetry of the electric field gradient
    with respect to static magnetic field.
    The use of the Rabi-frequency for EFG determination, including asymmetry,
    has been studied in the case of pure NQR
    through the use of nutation spectroscopy~\cite{sinjavsky1998, harbison1989}.
    Computational results of nutation spectroscopy in the presence of a magnetic field
    have also been performed \cite{spencer2013, spencer2011},
    but lack analytical expressions.
    This paper remedies this by presenting exact expressions of the Rabi frequencies
    for the case of EFG alignment with the magnetic field.  

    We take the Fe-based superconductors (FeSCs) as an example
    and apply our expressions to demonstrate how they can be used for analysis.
    These unconventional high-$ T_\text{c} $ superconductors
    are perfectly suited for the type of analysis described
    in this paper
    since they have spin-3/2 nuclei
    which experience a large $ \sim $1 T internally-supplied magnetic field
    from the antiferromagnetic ordering of Fe magnetic moments.
    The orthorhombic/tetragonal crystal symmetry
    dictates that the field is always oriented along
    one of the EFG principal axes \cite{mazin2010, carretta2020}.
    Therefore zero external-field NMR (ZNMR) and a powder sample can be used
    since both the field and EFG  are tied to the crystal symmetry.
    Detailed analysis of these ZNMR spectra has been limited
    due to the lack of such expressions that we bring to light in this paper.
    These materials are of
    immediate interest in superconductivity research
    and the broader condensed-matter field,
    because of the interplay of magnetic and electrical order~\cite{ansari2023},
    as well as controversies surrounding the superconductivity mechanism.
    Other applications can be found in vapor cells \cite{donley2009}
    and other antiferromagnetic materials \cite{narath1965, bennett2009}.

    So, in this paper, we add to the body of knowledge on spin-3/2 particles
    in magnetic fields by studying the case of the field oriented
    along one of the EFG principal axes.
    The paper is organized as follows.
    In the Theory section, we start by using spin 1/2 formalism
    to derive easy-to-use
    analytical expressions of the eigenenergies, eigenfrequencies, and eigenstates,
    along with the matrix elements of the spin angular momentum operators.
    Formulas for the expectation value of the nuclear magnetic moment
    and the Rabi frequencies are also presented for the first time.
    The Results section is split into three subsections:
    (1) the predicted Zeeman-quadrupolar spectrum is visualized
    and two case studies of existing FeSC ZNMR data are examined
    to demonstrate the analysis using the derived expressions;
    (2) the full solution of the continuum,
    that is for all possible magnetic field and EFG strengths is shown;
    and (3) the geometric behavior of the Rabi frequencies is discussed.

    \section{Theory}
    An electrically quadrupolar nucleus which has a quadrupole moment $ Q $
    and a nuclear magnetic moment,
    in the presence of an EFG and a magnetic field,
    will experience energy level splitting due to both of these interactions.
    The EFG comes from the local electronic environment
    and is a tensor defined as
    $ V_{ij} = \partial^2 V/\partial x_i \partial x_j|_{\mathbf{r}_0} $
    where $ V $ is the electric potential, 
    and is evaluated at the position of the nucleus $ \mathbf{r}_0 $.
    A set of three principal axes can be determined upon diagonalization
    of the tensor.
    With one axis of the EFG principal axes
    assumed to align with the magnetic field in the $z$-direction,
    the dominant Hamiltonian
    is given by the sum of the Zeeman Hamiltonian and the quadrupolar Hamiltonian
    \cite{slichter1990}:
    \begin{equation}
        H = \hbar \left[I_z \omega_0 
        + \frac{1}{6} \omega_Q (3 I_z^2 - I^2)
        + \frac{1}{12} \omega_\perp (I_+^2 + I_-^2) \right],
        \label{eq:hamiltonian}
    \end{equation}
    where $ \omega_0 $ is the Zeeman frequency
    and $ \omega_Q $ is the quadrupole coupling constant defined as
    \begin{equation}
        \omega_0 = -\gamma B_0,
        \label{eq:w0}
    \end{equation}
    \begin{equation}
        \omega_Q = \frac{3eQ}{2I(2I - 1) \hbar} V_{zz}.
    \end{equation}
    And $ \omega_\perp $ is defined so that
    $ \omega_\perp/\omega_Q = (V_{xx} - V_{yy})/V_{zz} \equiv \eta $
    (commonly referred to as the EFG asymmetry parameter):
    \begin{equation}
        \omega_\perp = \frac{3eQ}{2I(2I - 1) \hbar} (V_{xx} - V_{yy}).
    \end{equation}
    Note that $ \eta $ is unrestricted in its range
    since $ V_{zz} $ is not necessarily the largest EFG component.
    The static magnetic field is denoted by $ B_0 $,
    $ e $ is the elementary charge, $ \hbar $ is the reduced Planck constant,
    $ \gamma $ is the nuclear gyromagnetic ratio,
    and $ \mathbf{I} $ is the nuclear spin angular momentum operator
    divided by $ \hbar $.

\begin{figure}[htbp]
        \centering
        \includegraphics{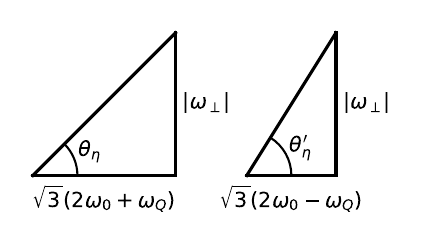}
        \caption{%
            Visual definitions of $ \theta_\eta $ and $ \theta_\eta' $,
            also defined in Eqs.~\ref{eq:theta_eta}-\ref{eq:theta_etap}.
            These angles can serve as a metric
            of transverse field asymmetry
            compared to longitudinal field components.%
        }
        \label{fig:triangle}
    \end{figure}

\begin{figure}[htbp]
        \centering
        \includegraphics[scale=1]{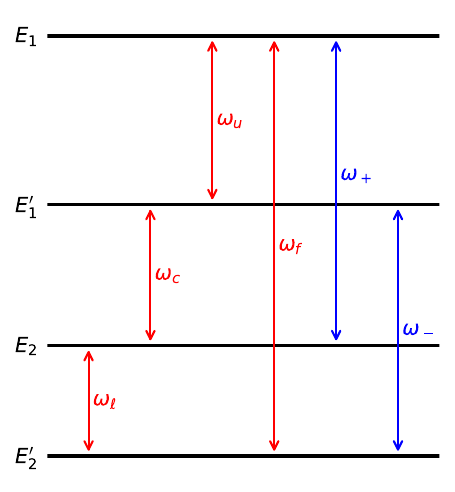}
        \caption{%
            Energy levels of the combined
            Zeeman-quadrupolar Hamiltonian.
            The spacings are drawn to scale,
            having been calculated for a system with
            $ R = 0.81 $ and $ \theta_\eta = 0.1 $.
            The transition frequencies are also labeled
            and defined in Table~\ref{tab:spectra}.
            The O-transitions are shown in red, and the P-transitions in blue,
            also defined in Table~\ref{tab:spectra}.%
        }
        \label{fig:energy}
    \end{figure}

    The Hamiltonian in Eq.~\ref{eq:hamiltonian}
    can be diagonalized in a number of ways
    but is perhaps more elegantly done
    by using fictitious spin-1/2 formalism.
    Doing so
    results in the eigenenergies and eigenstates (in the eigenbasis of $ I_z $)
    as given in Table~\ref{tab:eigen}
    (see Compliment B$_\text{IV}$ and C$_\text{IV}$
    of \cite{cohen-tannoudji2020}).
    As shown in Table~\ref{tab:eigen},
    they are defined in terms of the quantities
    $ \theta_\eta $, $ \theta_\eta' $, and $ \omega_\pm $
    which are given by
    \begin{equation}
        \tan\theta_\eta =
        \frac{|\omega_\perp|}{\sqrt{3} (2\omega_0 + \omega_Q)}
        \label{eq:theta_eta}
    \end{equation}
    \begin{equation}
        \tan\theta'_\eta =
        \frac{|\omega_\perp|}{\sqrt{3} (2\omega_0 - \omega_Q)},
        \label{eq:theta_etap}
    \end{equation}
    \begin{equation}
        \omega_\pm 
        =
        \sqrt{\left(2 \omega_0 \pm \omega_Q\right)^2 + \frac{\omega_\perp^2}{3}}.
        \label{eq:wpm}
    \end{equation}
    The angles are limited such that
    $ 0 \leq \theta_\eta, \theta'_\eta < \pi $.
    A geometric visualization of the angles is given in Fig.~\ref{fig:triangle}.
    Upon inspection, the system can be characterized by just three parameters:
    $ \omega_+ $, revealing the net strength of the combined field components;
    $ \theta_\eta $, a metric of the EFG asymmetry;
    and the relative ratio of the Zeeman and quadrupole contributions,
    \begin{equation}
        R = \frac{2\omega_0 - \omega_Q}{2\omega_0 + \omega_Q}.
    \end{equation}
    An energy level diagram
    calculated from these quantities
    is shown in Fig.~\ref{fig:energy}.
    \begin{table}[htbp]
        \centering
        \renewcommand{\arraystretch}{2.5}
        \caption{%
            Eigenstates and eigenenergies
            of the combined Zeeman-quadrupolar Hamiltonian
            given in Eq.~\ref{eq:hamiltonian},
            with angles and frequencies given in
            Eqs.~\ref{eq:w0}, \ref{eq:theta_eta}-\ref{eq:wpm}.%
        }
        \label{tab:eigen}
        \begin{ruledtabular}
        \begin{tabular}{@{}lrr@{}}
            Label
            & \multicolumn{1}{c}{Eigenenergy}
            & \multicolumn{1}{c}{Eigenstate}
            \\ \midrule
            1
            & $ \displaystyle \frac{\hbar}{2} (\omega_0 + \omega_+) $
            & $ \displaystyle \cos\frac{\theta_\eta}{2} \left|\frac{3}{2}\right\rangle + \sin\frac{\theta_\eta}{2}\left|\frac{-1}{2}\right\rangle $
            \\
            2
            & $ \displaystyle \frac{\hbar}{2} (\omega_0 - \omega_+) $
            & $ \displaystyle -\sin\frac{\theta_\eta}{2} \left|\frac{3}{2}\right\rangle + \cos\frac{\theta_\eta}{2}\left|\frac{-1}{2}\right\rangle $
            \\
            $ 1' $
            & $ \displaystyle \frac{\hbar}{2} (-\omega_0 + \omega_-) $
            & $ \displaystyle \cos\frac{\theta_\eta'}{2} \left|\frac{1}{2}\right\rangle + \sin\frac{\theta_\eta'}{2}\left|\frac{-3}{2}\right\rangle $
            \\
            $ 2' $
            & $ \displaystyle \frac{\hbar}{2} (-\omega_0 - \omega_-) $
            & $ \displaystyle  -\sin\frac{\theta_\eta'}{2} \left|\frac{1}{2}\right\rangle + \cos\frac{\theta_\eta'}{2}\left|\frac{-3}{2}\right\rangle $
        \end{tabular}
        \end{ruledtabular}
    \end{table}
    
    From these energy levels, six possible transition frequencies
    can be observed as shown in  Fig.~\ref{fig:energy},
    and are given in Table~\ref{tab:spectra}.
    The expressions for the eigenfrequencies are in agreement
    with previous work using low-field perturbation theory
    and exact expressions in the EFG-aligned case \cite{brooker1974}.
    As discussed in more detail below,
    the observability of these transitions
    depends on the radiofrequency excitation direction
    with respect to the magnetic field.
    More specifically, four of these transitions are completely unresponsive
    to excitations parallel to the magnetic field
    and are labelled as the ``O'' transitions (for ``orthogonal'').
    The other two transitions are unresponsive
    to excitations orthogonal to the magnetic field
    and are labelled as the ``P'' transitions (for ``parallel'').
    The P-transitions are equal
    to $ \omega_\pm $ defined in Eq.~\ref{eq:wpm}.
    \begin{table*}[htbp]
        \centering
        \renewcommand{\arraystretch}{1.75}
        \caption{%
            Exact expressions for the six possible transition frequencies
            and their associated spin angular momentum operator matrix elements.
            The first four lines of the table are those transitions
            which are unresponsive to excitations parallel to the magnetic field,
            labeled as the ``O'' transitions,
            and are given analogous names familiar in the high-field NMR limit.
            The last two unnamed transitions are unresponsive to excitations
            orthogonal to the magnetic field and are labeled as the ``P'' transitions.
            The labeling of the transitions (O vs.\ P)
            is with respect to maximum responsiveness.
            The angles and frequencies are given in
            Eqs.~\ref{eq:w0}, \ref{eq:theta_eta}-\ref{eq:wpm}.%
        }
        \label{tab:spectra}
        \begin{ruledtabular}
        \begin{tabular}{@{}lrrrrl@{}}
            $ p\phantom{'} \leftrightarrow q\phantom{'} $
            & \multicolumn{1}{c}{Frequency}
            & \multicolumn{1}{c}{$ \left\langle p \right| I_+ \left| q \right\rangle $}
            & \multicolumn{1}{c}{$ \left\langle p \right| I_- \left| q \right\rangle $}
            & \multicolumn{1}{c}{$ \left\langle p \right| I_z \left| q \right\rangle $}
            & Transition Name
            \\ \midrule
            \multicolumn{6}{@{}l@{}}{Orthogonal ``O'' Transitions} \\
            $ 2\phantom{'} \leftrightarrow 2' $
            & $ \omega_0 - \frac{\omega_+}{2} + \frac{\omega_-}{2} $
            & $ \sqrt{3} \cos\left( \frac{\theta_\eta - \theta'_\eta}{2} \right) $
            & $ -2 \cos\frac{\theta_\eta}{2} \sin\frac{\theta'_\eta}{2} $
            & \multicolumn{1}{c}{0}
            & $ \omega_\ell $, lower satellite
            \\
            $ 1' \leftrightarrow 2\phantom{'} $
            & $ -\omega_0 + \frac{\omega_+}{2} + \frac{\omega_-}{2} $
            & $ 2 \cos\frac{\theta_\eta}{2} \cos\frac{\theta'_\eta}{2} $
            & $ -\sqrt{3} \sin\left( \frac{\theta_\eta - \theta'_\eta}{2} \right) $
            & \multicolumn{1}{c}{0}
            & $ \omega_\text{c} $, central
            \\
            $ 1\phantom{'} \leftrightarrow 1' $
            & $ \omega_0 + \frac{\omega_+}{2} - \frac{\omega_-}{2} $
            & $ \sqrt{3} \cos\left( \frac{\theta_\eta - \theta'_\eta}{2} \right) $
            & $ 2 \cos\frac{\theta'_\eta}{2} \sin\frac{\theta_\eta}{2} $
            & \multicolumn{1}{c}{0}
            & $ \omega_\text{u} $, upper satellite
            \\
            $ 1\phantom{'} \leftrightarrow 2' $
            & $ \omega_0 + \frac{\omega_+}{2} + \frac{\omega_-}{2} $
            & $ \sqrt{3} \sin\left( \frac{\theta_\eta - \theta'_\eta}{2} \right) $
            & $ -2 \sin\frac{\theta_\eta}{2} \sin\frac{\theta'_\eta}{2} $
            & \multicolumn{1}{c}{0}
            & $ \omega_\text{f} $, forbidden
            \\ \midrule
            \multicolumn{6}{@{}l@{}}{Parallel ``P'' Transitions} \\
            $ 1\phantom{'} \leftrightarrow 2\phantom{'} $
            & \multicolumn{1}{c}{$ \omega_+ $}
            & \multicolumn{1}{c}{0}
            & \multicolumn{1}{c}{0}
            & $ -\sin\theta_\eta $
            \\
            $ 1' \leftrightarrow 2' $
            & \multicolumn{1}{c}{$ \omega_- $}
            & \multicolumn{1}{c}{0}
            & \multicolumn{1}{c}{0}
            & $ -\sin\theta'_\eta $
        \end{tabular}
        \end{ruledtabular}
    \end{table*}

    A measurement is performed by sending a time-dependent perturbation,
    a radio-frequency pulse with frequency $ \omega $ resonant with one of the six transitions,
    \begin{equation}
        H_1 = \hbar \omega_1 \mathbf{I} \cdot \hat{n} \cos(\omega t),
    \end{equation}
    where $ \omega_1 = -\gamma B_1 $ is the Rabi frequency
    and $ B_1 $ is the amplitude of the signal.
    The excitation direction is
    $ \hat{n} = \sin\theta\cos\varphi\hat{x}
    + \sin\theta\sin\varphi\hat{y} + \cos\theta\hat{z} $,
    where $ \varphi $ and $ \theta $ are the usual azimuth and polar angles
    with respect to $ \mathbf{B}_0 $.
    The response is a net nuclear magnetic moment.
    For a transition from $ p $ to $ q $, its expectation value
    can be calculated using fictitious spin-1/2 formalism
    and is given by
    \begin{equation}
        \langle \boldsymbol{\mu} \rangle_{pq} =
        i \gamma \hbar f_{pq}  
        \sin (\lambda_{pq} \Theta)
        \langle p | \mathbf{I} | q \rangle
        \mathrm{e}^{-i (\omega_{pq} t - \xi_{pq})} + \mathrm{c.c.},
        \label{eq:signal}
    \end{equation}
    where c.c.\ stands for the complex conjugate of the preceding term,
    $ f_{pq} = \hbar \omega_{pq}/[2 (2I + 1) k_B T] $
    is the Boltzmann factor,
    the tip angle is $ \Theta = \omega_1 t_\text{p} $,
    where $ t_\text{p} $ is the pulse time,
    the transition frequency is denoted by $ \omega_{pq} $,
    and $ \xi_{pq} $ is the argument
    of $ \langle p |\mathbf{I} \cdot \hat{n}| q \rangle $.
    The Rabi frequency is modified by an efficiency coefficient
    \begin{equation}
        \lambda_{pq} \equiv |\langle p |\mathbf{I} \cdot \hat{n} | q \rangle|,
        \label{eq:lambda}
    \end{equation}
    a unitless factor.
    In practice, when excitation is in the same direction as detection,
    Eq.~\ref{eq:signal} shows that the signal
    is proportional to $ \lambda_{pq} \sin(\lambda_{pq}\Theta) $,
    and is therefore proportional to $ \lambda_{pq}^2 $
    in the small angle limit.
    The utility of the efficiency coefficient is that it contains
    all the geometric information of the effective excitation.
    Using Table~\ref{tab:spectra}, the efficiency coefficients
    are a straightforward calculation.
    For the P-transitions, since $ I_+ $ and $ I_- $ are all zero,
    the coefficients are
    \begin{gather}
        \lambda_{12} = \sin\theta_\eta \cos\theta ,
        \label{eq:lambda_par1}
        \\
        \lambda_{1'2'} = \sin\theta'_\eta \cos\theta .
        \label{eq:lambda_par2}
    \end{gather}
    Note that these are directly proportional to the EFG asymmetry
    and therefore the parallel transitions can be a direct metric of the EFG asymmetry.
    And for any O-transitions they are
    \begin{align}
        \lambda^2_{pq} = \frac{1}{4} \sin^2\theta 
        &\left[
            \langle p | I_+ | q \rangle^2
            + \langle p | I_- | q \rangle^2 \right.
            \nonumber \\
            &\left. + 2 \langle p | I_+ | q \rangle \langle p | I_- | q \rangle \cos(2\varphi)
        \right].
        \label{eq:lambda_ortho}
    \end{align}
    Eqs.~\ref{eq:lambda_par1}-\ref{eq:lambda_ortho} are in agreement
    with expressions derived in the zero-field case \cite{odin1999}.

    \section{Results}
    Here,
    results stemming from the theory in the previous section
    are presented.
    We first show a simulated spectrum produced
    from a specific EFG and magnetic field strength
    and elucidate how it relates to previous experimental ZNMR data on FeSCs.
    Several relationships are highlighted that can be used to understand
    such previously unanalyzed data.
    The analyzed data and comparisons to other experiments are summarized
    in Table~\ref{tab:params}.
    Afterwards, we discuss the more general continuum of results,
    that is for any arbitrary EFG and magnetic field strength.
    We also discuss the ``detectability'' of the peaks,
    which will bring to light which parts of the continuum
    are easy or difficult to detect in an experiment.
    Lastly, we visualize the Rabi frequencies
    which are derived in Eqs.~\ref{eq:lambda_par1}-\ref{eq:lambda_ortho}.

    \subsection{Predicted Spectrum and Fe-based Superconductors Example}
\begin{figure*}[htbp]
        \centering
        \includegraphics{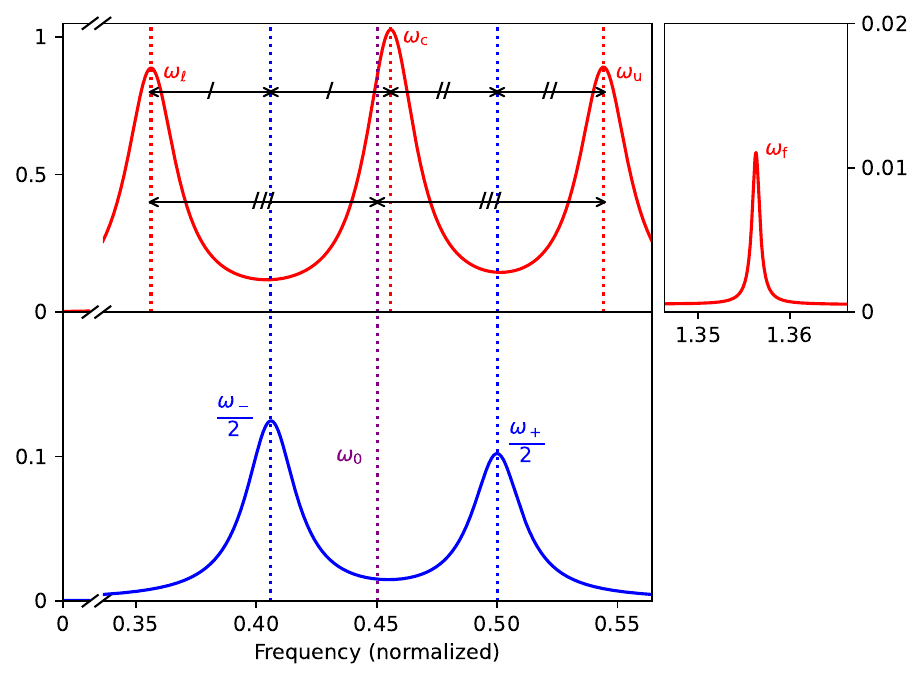}%
        \caption{%
            A simulated spectrum of the combined
            Zeeman-quadrupolar Hamiltonian
            made with $ R = 0.81 $ and $ \theta_\eta = 0.1 $,
            showing all 6 possible transitions
            and the Zeeman frequency $ \omega_0 $.
            All frequencies are normalized with respect to $ \omega_+ $.
            The O-transitions are shown in red,
            and the P-transitions in blue,
            as defined in Table~\ref{tab:spectra}.
            Peak widths were chosen arbitrarily.
            The frequencies and their values
            relative to neighboring transitions
            are illustrated by the black arrow lines.%
        }
        \label{fig:sample_spectrum}
    \end{figure*}

    The spectrum is simulated using the quantities 
    given in Table~\ref{tab:spectra}
    and is shown in Fig.~\ref{fig:sample_spectrum}.
    This spectrum was generated by choosing specific
    $ R $ and $ \theta_\eta $ values which
    closely resemble an experimental spectrum.
    All frequencies are normalized
    with  respect to $ \omega_+ $.
    The peak heights are calculated,
    using Eqs.\ref{eq:lambda_par1}-\ref{eq:lambda_ortho}, by
    $
        \sqrt{(\langle p|I_+|q\rangle^2 + \langle p|I_-|q\rangle^2)/4
              + \langle p|I_z|q\rangle^2}
    $ and then normalized with respect to the largest peak.
    Finally, artificial broadening was applied to enhance visibility.
    The O/P-transitions,
    defined in Table~\ref{tab:spectra},
    are shown in the top/bottom panels and colored red/blue.
    In the figure, the O-transitions
    have been labeled by their first letters
    (e.g., $ \omega_\text{c} $ corresponds to the ``central'' transition);
    the P-transitions are simply labeled by $ \omega_\pm $.

    The predicted spectrum in Fig.~\ref{fig:sample_spectrum}
    was constructed to match that of particular FeSCs,
    which we have taken as a case study.
    FeSCs are characterized by their strong internal magnetic field
    which results from the antiferromagnetic ordering of Fe magnetic moments.
    The magnetic ordering is an intrinsic feature of these materials,
    and therefore both the magnetic field and the EFG
    are fixed to the crystal structure.
    Since the crystal structure of these materials
    is either orthorhombic or tetragonal, depending on the phase,
    symmetry dictates that the internal magnetic field will always be aligned
    with one of the EFG principal axes.
    For these reasons, FeSCs can be studied by ZNMR using
    either a powder sample or a single crystal.
    
    Several features in Fig.~\ref{fig:sample_spectrum} can be discussed.
    First, the Zeeman frequency is simply the average of
    the lower and upper satellites.
    Second, from the formula for the central transition,
    it is clear that if there is no EFG asymmetry,
    then $ \omega_\text{c} = \omega_0 $
    and the central transition would be exactly centered between the satellite peaks.
    Therefore asymmetry of the central peak with respect to the satellite peaks
    is a clear indication of EFG asymmetry.
    Lastly, the quadrupole coupling constant can be determined by
    subtraction in quadrature of adjacent peak averages.
    With regards to the P-transitions,
    it is more fruitful to discuss them in terms of
    half of their frequency values,
    which is how they are plotted in Fig.~\ref{fig:sample_spectrum}.
    This is because the average of the lower and central transitions
    is $ \omega_-/2 $.
    Similarly, 
    the average of the central and upper transitions is $ \omega_+/2 $.
    Finally, the forbidden transition is the sum of the
    lower, central, and upper transitions,
    which is more evidently displayed in Fig.~\ref{fig:energy}.
    With these relationships,
    one only needs to know three transitions
    in order to determine the other three.
    
\begin{figure*}[htbp]
        \centering
        \includegraphics[scale=1]{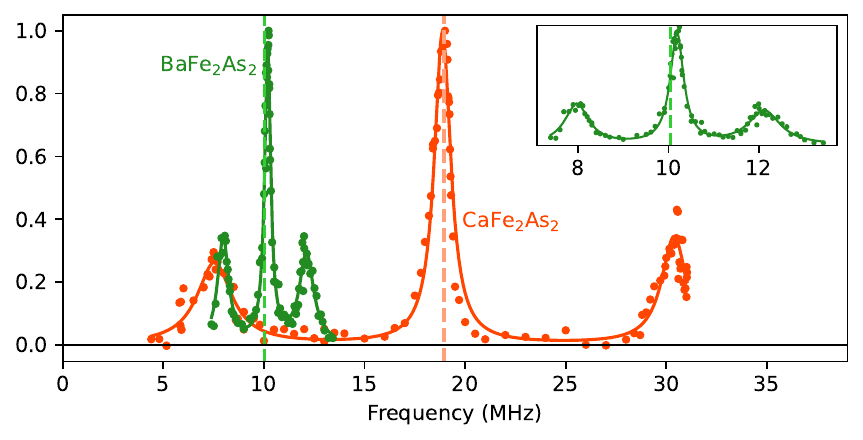}
        \caption{%
            The experimental spectra of
            BaFe$_2$As$_2$ \cite{fukazawa2008} and CaFe$_2$As$_2$ \cite{kawasaki2011},
            both performed at low-temperature
            and at zero external-field with powder samples.
            Data points were taken from literature and fitted with 3 Lorentzian curves.
            The vertical dashed lines represent the Zeeman frequency for each system.
            These two systems are from the same family of FeSCs
            yet they represent two extremes of the EFG,
            indicated by central peak deviation from the Zeeman frequency,
            with BaFe$_2$As$_2$ having high EFG asymmetry
            and CaFe$_2$As$_2$ having zero EFG asymmetry.
            Larger EFG-quadrupolar coupling is indicated by larger
            spread of the spectra.
            The quadrupole coupling constant, EFG asymmetry parameter,
            and internal field values are given in Table~\ref{tab:params}.%
        }
        \label{fig:exp_spectra}
    \end{figure*}

    The above analysis can be illustrated
    by examining two structurally similar FeSCs
    with very different spectra:
    BaFe$_2$As$_2$ \cite{fukazawa2008}
    and CaFe$_2$As$_2$ \cite{kawasaki2011},
    which represent all available $ ^{75}$As ZNMR data
    with 3 peaks on FeSCs, to date.
    Both of these materials were studied using a powder sample.
    We fit the existing experimental data,
    as shown in Fig.~\ref{fig:exp_spectra}.
    Using both the fits to determine the frequencies
    and the equations
    of $ \omega_\ell $, $ \omega_\text{c} $, and $ \omega_\text{u} $
    found in Table~\ref{tab:spectra},
    we are able to solve for the quadrupole coupling constant,
    internal magnetic field, and EFG asymmetry parameter.
    These values are shown in Table~\ref{tab:params}
    along with other literature values for comparison.
    The values in that table agree with
    some observations of Fig.~\ref{fig:exp_spectra}.
    Namely, one can immediately see that the internal field
    of CaFe$_2$As$_2$ is nearly double that of BaFe$_2$As$_2$,
    as determined by the average of the satellite peaks shown as a dashed line
    in the figure.
    Further, the quadrupole coupling constant of CaFe$_2$As$_2$
    is 6 times larger than that of BaFe$_2$As$_2$,
    as evidenced by the spread of the spectra around the central peak.
    The EFG asymmetry,
    which is indicated by the shift of the central peak
    from the Zeeman frequency,
    is clearly evident in BaFe$_2$As$_2$,
    but not in CaFe$_2$As$_2$.
    Finally, it should be noted that
    due to the small values of $ \theta_\eta $ for these materials,
    the forbidden and P-transitions would be difficult to directly detect
    and, as expected, were not reported in the literature.
    In addition, $ \eta $ values were not reported
    perhaps due to the difficulty of extracting it
    without the use of analytic equations.
    Instead, we extracted those values by fitting the data
    reported in the literature, as described above.
    \begin{table*}[htbp]
        \centering
        \caption{%
            Extracted parameters of BaFe$_2$As$_2$ and CaFe$_2$As$_2$.
            Rows with an asterisk were determined by solving
            the frequency equations from Table~\ref{tab:spectra}
            with the fitted peaks from Fig.~\ref{fig:exp_spectra}
            for the 3 parameters;
            otherwise they are quoted directly from literature.
            We use the standard notation:
            $ \nu_Q $ is the quadrupole coupling constant,
            $ H_\mathrm{int} $ is the internal magnetic field,
            and $ \eta $ is the EFG asymmetry parameter
            defined by $ |V_{aa} - V_{bb}|/|V_{cc}| $,
            where $ a $, $ b $, and $ c $ are the lattice parameters
            of the crystal structure.
            SC refers to ``single crystal''
            and $ H $ is an externally applied magnetic field.
        }
        \label{tab:params}
        \renewcommand{\arraystretch}{1.3}
        \begin{ruledtabular}
        \begin{tabular}{@{}lccccc@{}}
            Material & $ \nu_Q $ (MHz) & $ H_\mathrm{int} $ (T) & $ \eta $ & Methodology & Temperature (K)
            \\ \hline
            CaFe$_2$As$_2$ \cite{baek2009}
            & $ 12.4(1) $ & 2.6(1) & - & NMR SC $ H || c $ & 20
            \\
            CaFe$_2$As$_2$ \cite{kawasaki2011}$^\star$
            & $ 11.4(1) $ & 2.59(1) & 0.0(1) & ZNMR powder & 5
            \\
            CaFe$_2$As$_2$ \cite{dioguardi2012}
            & $ 12.38 $ & 2.628 & 0.39(5) & NMR SC rotation & 20 
            \\
            CaFe$_2$As$_2$ \cite{cui2015}
            & $ 12.9(2) $ & 2.6(1) & - & NMR SC $ H || c $ & 4.3 
            \\
            BaFe$_2$As$_2$ \cite{kitagawa2008}
            & $ 2.21(1) $ & 1.46(4)  & 1.18(3) & NMR SC rotation & 7 
            \\
            BaFe$_2$As$_2$ \cite{fukazawa2008}$^\star$
            & $ 2.09(3) $ & 1.377(4) & 2.0(2) & ZNMR powder & 1.5 
        \end{tabular}
        \end{ruledtabular}
    \end{table*}

    In addition to the parameters we extracted,
    Table~\ref{tab:params} also contains the same parameters,
    as quoted in the literature from high-field NMR experiments
    that used a single crystal sample.
    Comparing between different methodologies for the same material,
    the quadrupole coupling constant and the internal field
    are in good agreement with each other.
    However, the EFG asymmetry parameter has notable disagreements
    for CaFe$_2$As$_2$,
    with ZNMR finding no EFG asymmetry but high-field NMR
    finding a sizeable value of 0.39(5).
    It is unclear why this is the case,
    but it has been shown that the structure in this material is sensitive
    to different growth methodologies \cite{Saparov2014, furukawa2014}.
    We suspect that depending on how the sample is grown,
    strain may exist that contributes to the EFG.

    \subsection{Full Solution Detectability}
\begin{figure*}[htbp]
        \centering
        \includegraphics[scale=1]{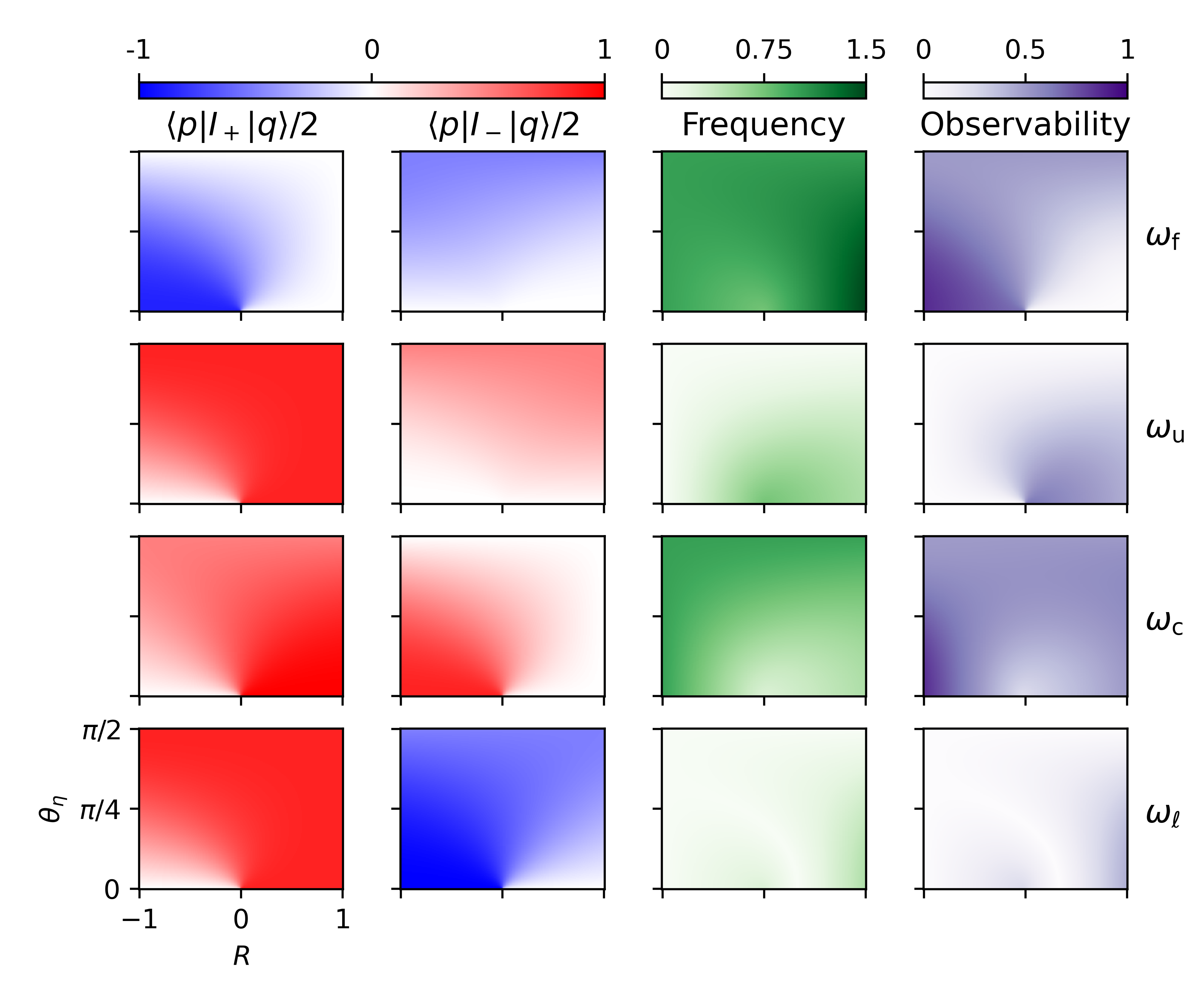}
        \caption{%
            Full solution of the Zeeman plus quadrupolar Hamiltonian spectrum
            for the O-transitions,
            as defined in Table~\ref{tab:spectra},
            as a function of $ R $ and $ \theta_\eta $.
            Each row corresponds to a different transition frequency,
            as labeled on the right side of the figure.
            Each column represents a different quantity of interest
            as labeled on the top of the figure.
            The first two columns are the matrix elements
            of the raising and lowering operators.
            Here, $ p $ and $ q $ refer to the labels of the eigenstates
            involved in the transition.
            The third column represents the actual transition frequency.
            Finally, the fourth column is the observability,
            which is calculated as
            $ \omega_{pq}\sqrt{\langle p|I_+|q\rangle^2 + \langle p|I_-|q\rangle^2}/2 $,
            where $ \omega_{pq} $ is the transition frequency of the row.
            All frequency quantities are normalized with respect to $ \omega_+ $.
        }
        \label{fig:ortho}
    \end{figure*}

\begin{figure*}[htbp]
        \centering
        \includegraphics[scale=1]{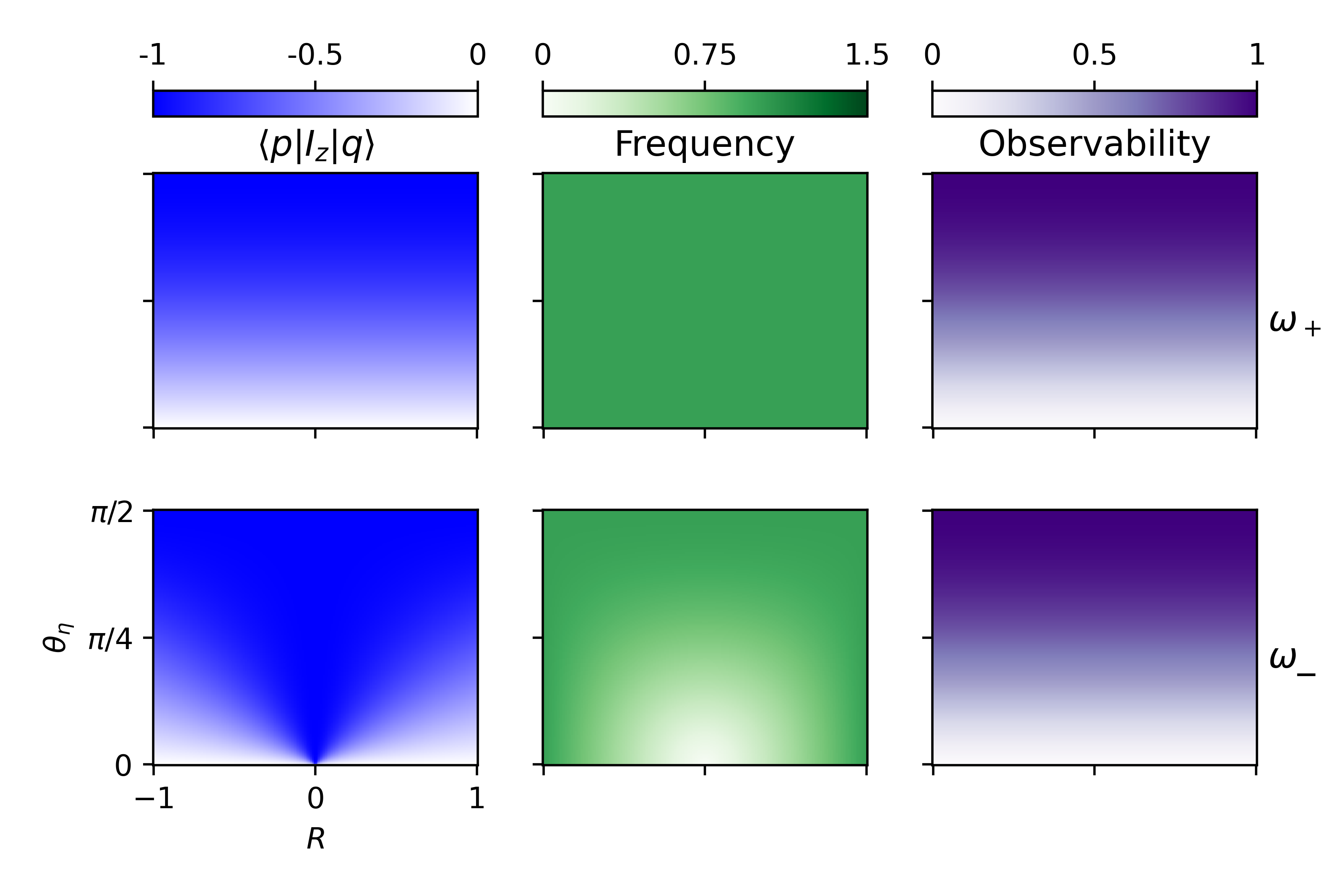}
        \caption{%
            Similar to Fig.~\ref{fig:ortho}, except the two rows correspond
            to the P-transitions, as defined in Table~\ref{tab:spectra}.
            The observability is calculated by
            $ \omega_{pq} |\langle p | I_z | q \rangle| $,
            where $ \omega_{pq} $ is the transition frequency of the row.
            All frequency quantities are normalized with respect to $ \omega_+ $.%
        }
        \label{fig:par}
    \end{figure*}

    The two case studies in the previous section
    only represent two sets of $ (R, \theta_\eta) $ values.
    Depending on the strength of the magnetic field and the EFG,
    spectra could be substantially different.
    Therefore, we show visualizations
    of the full solution set
    for the O-transitions in Fig.~\ref{fig:ortho}
    and the P-transitions in Fig.~\ref{fig:par}.
    From these, one can quickly gain a sense
    of which Zeeman-quadrupolar strengths and EFG asymmetry
    give frequencies that can be observed and from which direction.
    
    In the limit of small $ \theta_\eta $
    and $ R \to 1 $, which corresponds to high-field NMR,
    only three of the four O-transitions are observable:
    $ \omega_\text{u} $, $ \omega_\text{c} $, and $ \omega_\ell $.
    Also, as expected, they are responsive only to circularly polarized magnetic fields,
    co-rotating with the standard rotating frame,
    with  $ \langle p|I_-|q\rangle \to 0 $. 
    The $ \langle p|I_+|q\rangle $ ratio
    of the lower, central, and upper signals
    is found to be $ \sqrt{3} $:2:$ \sqrt{3} $,
    as has been found previously\cite{bain2003, ramsay2020}.
    
    On the other hand, when $ \omega_0 \to 0 $,  corresponding to the NQR limit,
    $ \theta'_\eta = \pi-\theta_\eta $ and $ R \to -1 $ for finite $ \omega_Q $.
    In this limit, there is only one observable frequency,
    indicated by the equal values of
    $ \omega_\text{f} $, $ \omega_\text{c} $, and $ \omega_\pm$, as expected.
    This is due to the double degeneracy of spin-3/2 nuclei in zero-field:
    $ E_1 = E_1' $ and $ E_2 = E_2' $.
    This creates two energy levels,
    each populated by nuclei with equal energies
    but mutually opposite spin states.
    A linearly polarized excitation will therefore excite
    both spin states,
    one for each circularly polarized component of the incident pulse.
    Hence the observed signal will also be linearly polarized,
    oscillating in the same direction as the excitation \cite{weber1960}.
    This is evident in the fact that
    $ \omega_\text{f} $ and $ \omega_\text{c} $
    together have an equal but opposite response
    to the different helicity excitation;
    that is, $ \langle 1|I_+|2' \rangle = - \langle 1'|I_-|2 \rangle $
    in this limit.
    Finally, since the signals of $ \omega_\pm $ are inherently linear,
    we can conclude that the overall response is linear.

    In the case of comparable Zeeman and quadrupolar strengths, $ -1 < R < 1 $.
    The observability of $ \omega_\text{u} $
    can be a metric of $ R $,
    particularly for lower asymmetry with $ \theta_\eta \leq \pi/4 $,
    since it disappears as $ R $ becomes increasingly negative.
    It could be useful for tracking an appearing/disappearing
    internal magnetic field close to the phase transitions,
    in the case of FeSCs for example.
    For general observability and tracking of dynamics,
    for instance $ T_1/T_2 $ data,
    the central line would be useful due to
    its strong responses for $ R $ not close to zero.
    
    For the P-transitions plot shown in Fig.~\ref{fig:par},
    only two rows are shown, corresponding to $ \omega_\pm $.
    Since all frequencies are normalized with respect to $ \omega_+ $,
    the frequency plot for the first P-transition
    which is defined as $ \omega_+ $
    is seen as constant.
    Main features of observation are that for small values of $ \theta_\eta $
    and all values of $ R $,
    the transitions would be difficult to observe directly.
    Detectability increases with EFG asymmetry,
    and may even serve as a metric of it
    as the Rabi frequency is directly proportional to $ \omega_\perp $.

    \subsection{Geometric Dependence of the Rabi Frequency}
    Finally, we examine the signal's dependence on the geometry of excitation and detection,
    by visualizing the Rabi frequency coefficient,
    as defined in Eq.~\ref{eq:lambda}.
    Assuming the use of a single crystal,
    we can use this coefficient to gain a better understanding
    of how the signal will behave for different geometric orientations.
    In particular, we are focused on signals
    such that excitation and detection
    are along the same direction with linear polarization.

\begin{figure*}[htbp]
        \centering
        \includegraphics[scale=0.5]{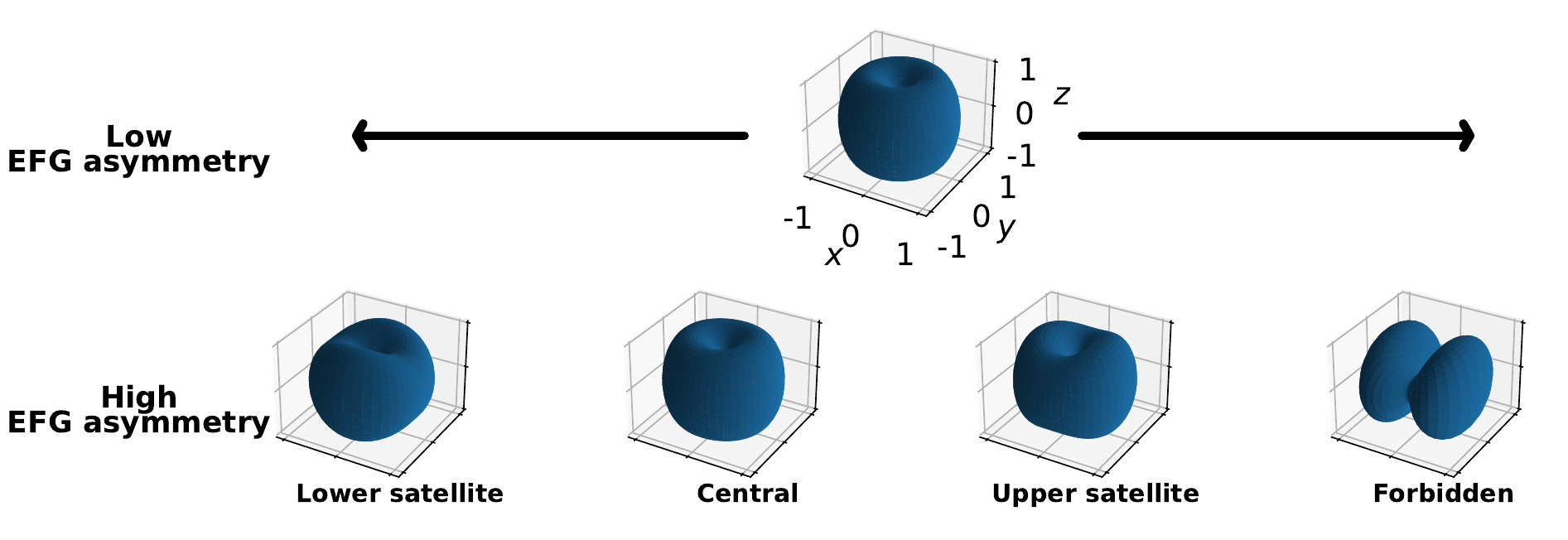}
        \caption{%
            Polar plots of the square of the Rabi frequency coefficients $\lambda^2$,
            the geometric modifier of the Rabi frequency,
            for the O-transitions,
            as defined in Eq.~\ref{eq:lambda_ortho}.
            These plots allow for a simple geometric representation
            of the observed signal,
            relative to the magnetic field which is oriented along $ z $.
            A particular value of $ R = 0.81 $ is chosen for all plots,
            but with $ \theta_\eta = 0.1 $
            for high EFG asymmetry (bottom row)
            and $ \theta_\eta = 0.001 $ for low EFG asymmetry (top row).
            Each column corresponds to a different O-transition,
            as labeled below each column.
            For the low EFG asymmetry row,
            all plots look nearly identical, hence only one plot is shown.
            All values have been normalized with respect to the maximum
            of each subplot.%
        }
        \label{fig:lambda_ortho}
    \end{figure*}

    Starting with the four O-transitions,
    their Rabi coefficients are given by Eq.~\ref{eq:lambda_ortho}.
    In particular, we plot $ \lambda^2_{pq} $ in 3D real space,
    as shown in Fig.~\ref{fig:lambda_ortho}.
    Due to the cyclic nature of the spin angular momentum operators
    in $ \theta_\eta $ and $ \theta_\eta' $,
    defined in Table~\ref{tab:spectra},
    all of the features of the plots can be
    reproduced by some phase shift.
    Namely, all of the Rabi coefficient plots are phase shifted from the central transition plots
    in the following ways:
    a $ \theta_\eta' - \pi $ phase shift for the lower satellite,
    a $ \theta_\eta + \pi $ phase shift for the upper satellite,
    and both a $ \theta_\eta' - \pi $ and $ \theta_\eta + \pi $ phase shift
    for the forbidden transition.
    For that reason, only the unique features of the plots
    are shown in the figure.
    For example, for high EFG asymmetry values,
    the signal can be highly asymmetric
    for some of the O-transitions.
    In contrast, for low EFG asymmetry,
    the O-transitions are all highly symmetric about the $ z $-axis.
    In all four plots however, the concept of the O-transition,
    that is transitions which are unaffected by excitations parallel
    to the magnetic field,
    is captured with obvious dips right on the $ z $-axis.

\begin{figure}[htbp]
        \centering
        \includegraphics{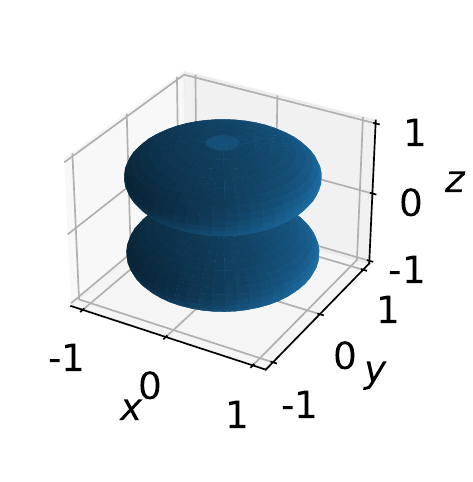}
        \caption{%
            This is also a plot of the Rabi frequency coefficient,
            similar to Fig.~\ref{fig:lambda_ortho},
            except for the P-transitions,
            defined in Eqs.~\ref{eq:lambda_par1}-\ref{eq:lambda_par2}.
            Only one transition is shown
            and for a single pair of $ (R, \theta_\eta ) $
            because the features of the plot are identical for all values.
            As expected, no signal is seen orthogonal to the $ z $ direction,
            which is parallel to the magnetic field.%
        }
        \label{fig:lambda_par}
    \end{figure}

    For the two P-transitions,
    the Rabi coefficients are shown in Fig.~\ref{fig:lambda_par}.
    Only one plot is shown because both plots, when normalized,
    have identical features for all values of $ R $ and $ \theta_\eta $.
    The plots are symmetric about $ z $,
    with the values collapsing to zero on the $x$-$y$ plane.
    This is sensible since the P-transitions
    are insensitive to excitations orthogonal to the magnetic field.
    This could be potentially useful with regards to materials
    with internal magnetic fields, such as the FeSCs.
    In this context, in theory with a single crystal,
    the geometry of the internal field and the EFG,
    could be determined with just the frequencies that are detected.

    \section{Conclusions}
    In this paper, the spectrum from the combined
    Zeeman and quadrupole Hamiltonian of spin-3/2 particles
    is calculated for the case where the magnetic field
    is aligned with one of the EFG principal axes.
    We derive succinct closed-form solutions for transition frequencies.
    In addition, for the first time,
    we derive magnetic moment arising from resonant excitation
    as function of excitation strength and direction
    as well as explicit expressions for the Rabi frequencies,
    given in Eqs.~\ref{eq:lambda_par1}-\ref{eq:lambda_ortho}.
    The results can be used not only for conventional NMR,
    where the magnetic field is applied externally,
    but also for systems which produce static internal magnetic fields.  

    An important example of materials with intrinsic fields
    can be found in FeSCs.
    Pre-existing $^{75}$As spectra from two such materials are analyzed.
    Because both the magnetic and EFG fields
    are intrinsic to the crystal's structure,
    these spectra, arising from powder samples, are not broadened.
    Therefore, the field values obtained compare well
    to single-crystal goniometer measurements done with conventional NMR.
    For these materials, three of the six possible transitions dominate.
    With analytic expressions,
    the spectra are readily interpreted%
    --the average of the satellite frequencies is the Zeeman frequency,
    subtraction in quadrature of adjacent peak averages
    is proportional to the quadrupole coupling constant,
    and the deviation of the central peak from the Zeeman frequency
    can be used to calculate the EFG asymmetry.
    While the first two calculations are straightforward,
    the latter is more involved.
    Therefore, the relative uncertainty on the asymmetry parameter
    is larger than on the other two field quantities.

    In contrast, the Rabi frequency of the P-transitions
    is found to be directly proportional to the EFG asymmetry parameter.  
    This dependency can potentially be exploited to characterize the asymmetry.
    For instance, excitation on one of the P-transitions
    will alter the populations of the corresponding energy levels
    shown in Fig.~\ref{fig:energy}, according to the strength of the associated Rabi frequency.
    The resulting population changes in the quantum levels can then be read-out
    on the more readily detectable O-transitions.
    Although the P-transitions are often weak,
    they are easily found once the three dominant O-transitions are known,
    since the P-transitions are themselves
    related to the frequencies of the O-transitions by averages,
    as shown in Fig.~\ref{fig:sample_spectrum}.
    There is precedence for such two-frequency schemes
    in other systems \cite{grechishkin1983, mozjoukhine2000, sauer2001}.
    Such indirect detection of the asymmetry parameter
    could be sensitive and therefore particularly useful for studying systems
    with small asymmetry.
    An interesting example is the use of asymmetry
    as a metric for spin-fluctuations in FeSCs \cite{ansari2023}.

    In conclusion, the exact predictions presented here
    provide a firm foundation for future research
    with spin-3/2 particles immersed in a magnetic field
    aligned with the EFG frame. 

    \begin{acknowledgments}
        We would like to thank Shinji Kawasaki
        for insightful conversations and for supplying his data to us for analysis.
        In addition, we would like to thank Igor Mazin and Nick Curro
        for their input.
        This work was supported by
        the National Science Foundation (Award No.\ 2214194)
        and the Quantum Science \& Engineering Center
        at George Mason University.
    \end{acknowledgments}

    \bibliography{main}

\end{document}